\documentclass[conference]{IEEEtran}

\input epsf
\usepackage{graphicx}

\begin{document}

\title{\LARGE Regression Test Suite for Payment Switch using jPOS}

\author{\authorblockN{A. Sardesai, H. Piplodwala, S. Hargunani, S. Sonawane, Prof. P. Joshi, V. Mohite}
\authorblockA{Computer Engineering Department, Pune Institute of Computer Technology}}

\maketitle

\begin{abstract}
The Payment Switch is an integral component of all modern payment and banking systems in India. The NPCI currently provides a simulator to test payment switches. However, this system has a few disadvantages viz. it lacks an API, it requires manual generation of each test case and during high server loads, the testing process may take a long time. Currently there aren’t any open source alternatives to the NPCI simulator. We propose a system which solves these shortcomings. Our proposed system simulates the NPCI system. It allows connection with switches that are to be tested and automates the process of generation and execution of test cases. It also has the capability to generate test reports and can be run locally.
\end{abstract}
\IEEEoverridecommandlockouts
\IEEEpeerreviewmaketitle

\section{Introduction}
Financial technology is rapidly advancing and with it, payments have
become synonymous with credit and debit card transactions. These transactions
are being fuelled by modern payment systems. The switch is an integral and
inseparable component of these systems. The NPCI (National Payments Corporation of India)
currently uses the ISO8583 platform to relay messages between payment switches.
The Switch generally offers a rules-based authorisation and switching solution that is merchant driven. Payment transactions are constantly routed between numerous acquirers and Payment
Service Providers. The switch is critical in the processing and validation of various
types of payments. 
\par Such an important component would naturally require frequent updates from
switch developers. Every update must be tested rigorously; otherwise it might
result in invalid transactions and immense losses for the switch provider.
Currently, evaluating these switches is a time-consuming and laborious process. It
entails hundreds of test cases being conducted by individual switch developers.
The NPCI has provided a simulator which allows testing of bank switches but it has the following problems:
\begin{enumerate} 
\item It offers a GUI which requires manual input of test cases from the tester.
\item It does not have an API which can allow automation of test case execution.
\item During high server loads or bad internet connectivity, the tests can be very slow to execute, increasing the overall testing time.
\item When the simulator is offline, testing cannot be performed.

\end{enumerate}

The ISO8583 protocol and building payment switches requires niche knowledge.
Hence, an open source implementation to solve this problem currently does not exist.

The jPOS library in Java is an open source library which allows us to create applications that deal with the ISO8583 protocol. We propose building a jPOS Q2 application that solves the aforementioned problems.

\section{Related Work}
\par In terms of work and computational resources, in \cite{article1} Prabhakar K. et al. discuss how re-testing is costly for maintenance. As a result, this study provides a comprehensive strategy to generating cases for regression testing, which predicts test work needed and noticing faults.

\par In \cite{article2} Prof. A. Ananda Rao and Kiran Kumar J delve deeper into cost-effective regression testing. The goal of this research is to figure out how to cut the cost of regression testing. In a black box environment, a method for reducing test suites for regression testing has been proposed. 

\par In \cite{article3} Susanne Rösch, Sebastian Ulewicz, Julien Provost and Birgit Vogel-Heuser examined and contrasted the various testing methodologies utilised in production automation. The purpose is to evaluate these approaches in terms of their applicability to the field of industrial automation in order to spot current trends and research needs. 

\par In \cite{inproceedings4} B. Korel, L.H. Tahat and B. Vaysburg proposed a model-based regression testing strategy that reduces regression test suites by using EFSM (Extended Finite State Machine) model dependence analysis. 

\par In \cite{10.1145/1295074.1295086} Yanping Chen, Robert L. Probert and Hasan Ural discuss extended dependence analysis. Based on EFSM dependence analysis, a model-based regression test suite (RTS) reduction strategy is suggested.

\section{Overview}
We propose building a jPOS Q2 application that solves the aforementioned problems. The general attributes of this application can be stated as follows:

\begin{enumerate}
    \item It simulates the NPCI switch. (national financial switch not NPCI)
    \item It is able to connect to other switches.
    \item It can send and receive ISO-Messages using various channels like ASCII, NAC and XML.
    \item Allows testers to define test case templates and automates test case generation.
    \item Generates a test report based on the execution of test cases after comparing the received response against the expected response.
    \item Allows users to interact with the system via a GUI or CLI.
    \item Can operate as a Server or Client based on requirements.
    \item Can be run locally and an external internet connection is not required.
\end{enumerate}

\begin{figure}
    \centering
    \includegraphics[scale=0.5]{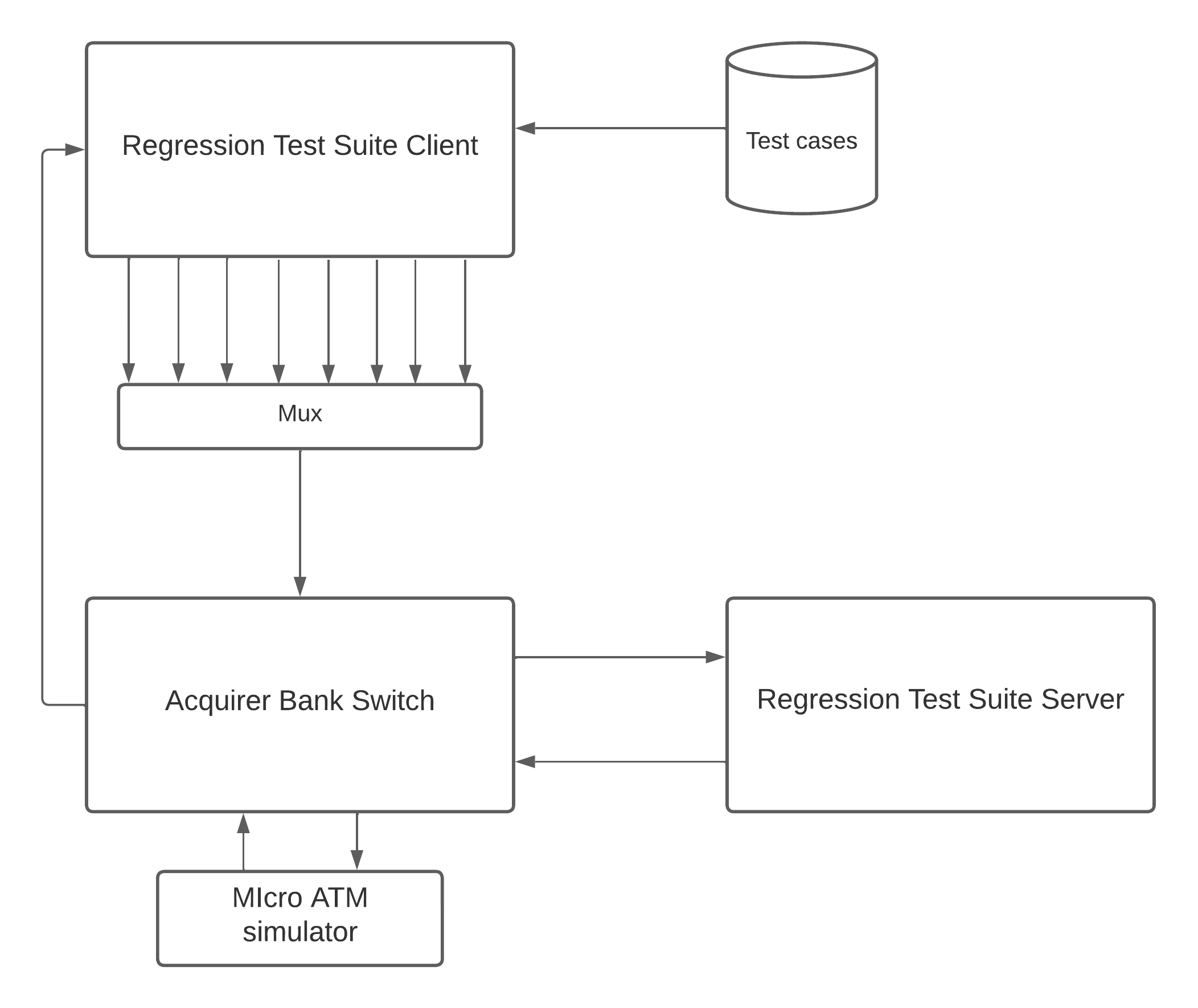}
    \caption{High Level Design}
    \label{fig:my_label}
\end{figure}

Further details are provided in the remainder of this paper for specific modules.

\section{System Modules}
We propose a system with the following key modules:
\subsection{ Message Generator}
Testers need to create a template for a particular test case. This is defined in a simple json file. The test case denotes an ISO8583 message. The randomize attribute allows us to randomize certain fields. 

\par The module is responsible for generating a message according to the ISO8583 specification. The message generator will use the json template to generate an ISOMsg instance while randomizing the necessary fields. A default field config file decides the regex for each field that needs to be randomized. This can be customized according to your custom packager.

\begin{figure}
    \centering
    \includegraphics[scale = 0.75]{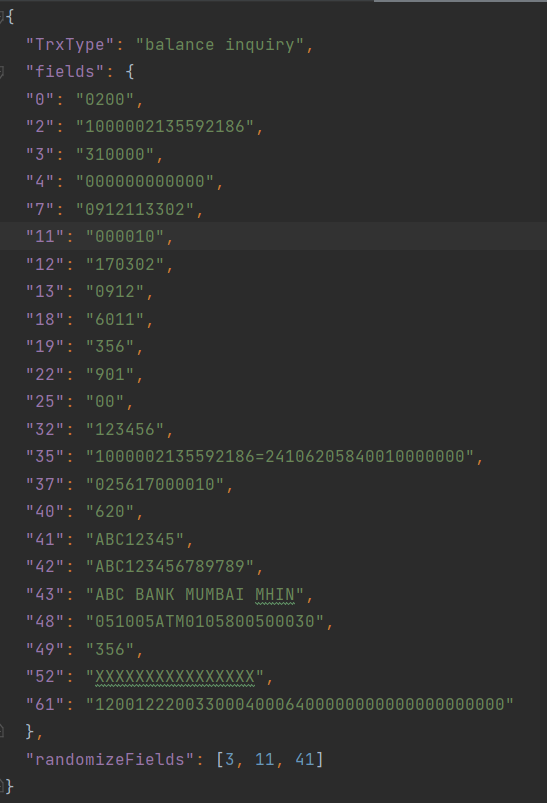}
    \caption{Message Generator - template for a sample transaction}
    \label{fig:my_label}
\end{figure}

\subsection{ Server }
The Server is a jpos Q2 application. It consists of three servers running on different ports. Each server caters to a specific ISO Channel (XML, ASCII, NAC). The transaction manager has a custom participant that validates the requests and returns an appropriate response. This is also where all of the business logic for request handling is implemented.\\
\par A function inside our custom participant to validate balance inquiry transactions:
\begin{flushleft}
\textbf{procedure} validateBalanceEnquiry() \newline
\hspace*{0.25in} 1 let response := "00" \newline\newline
\hspace*{0.25in} 2 if request[0] != "0200": \newline
\hspace*{0.5in} 2.1 response := "12"\newline\newline
\hspace*{0.25in} 3 if request[3] = "31[0-9]{4}": \newline
\hspace*{0.5in} 3.1 response := "12" \newline\newline
\hspace*{0.25in} 4 request.setMTI("0210") \newline\newline
\hspace*{0.25in} 5 request[39] := response \newline\newline
\hspace*{0.25in} 6 if response = "00" \newline
\hspace*{0.5in} 6.1 request[54] := balance\newline\newline
\hspace*{0.25in} 7 return request
\end{flushleft}

\subsection{ Client }
This is a jPOS Q2 application that is responsible for carrying out regression testing on the payment switch. The Client allows the user to select multiple channels to send transactions, as well as the number of message iterations. It asynchronously sends all the test transactions to the switch simultaneously and generates a report based on the switch's responses. The key functions performed by the client are: 
\begin{enumerate}
    \item Establish connection with switch
    \item Pack test transaction in ISO format
    \item Decode response and compare with expected output
    \item Generate report based on responses seen
\end{enumerate}

\begin{figure}
    \centering
    \includegraphics[scale=0.5]{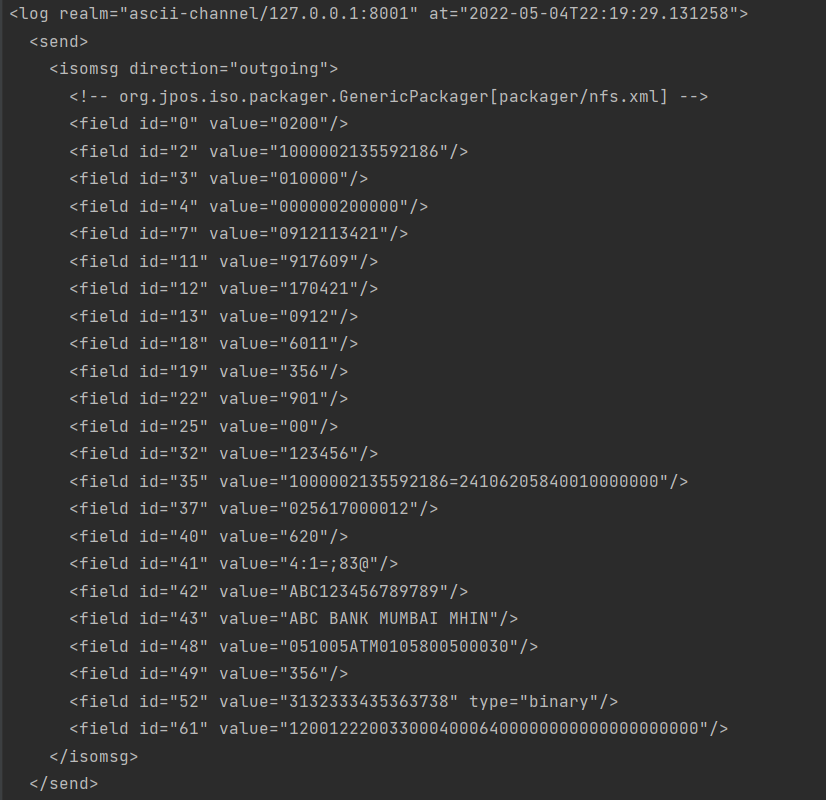}
    \caption{Client sending a transaction request to the server}
    \label{fig:my_label}
\end{figure}

\vspace*{0.1in}Fig. 3 shows the client sending a transaction request to the server, indicated by the isomsg direction value "outgoing". The request is sent through the ASCII channel on port 8001. The request body contains important information pertaining to the specific type of request, like terminal ID, MTI, bank branch, card number etc.

\subsection{ User Interface System }
We added a simple user interface system using Java Swing to ensure ease of usage for switch developers. This allows switch developers to perform regression testing on their system with just a few clicks. It also displays reports and allows them to be downloaded for future reference. The main features can be classified as:
\begin{enumerate}
    \item Allows selecting whether simulator should act as client or server
    \item Call client and server modules
    \item Display and allow downloading of reports
\end{enumerate}

\begin{figure}
    \centering
    \includegraphics[scale=0.5]{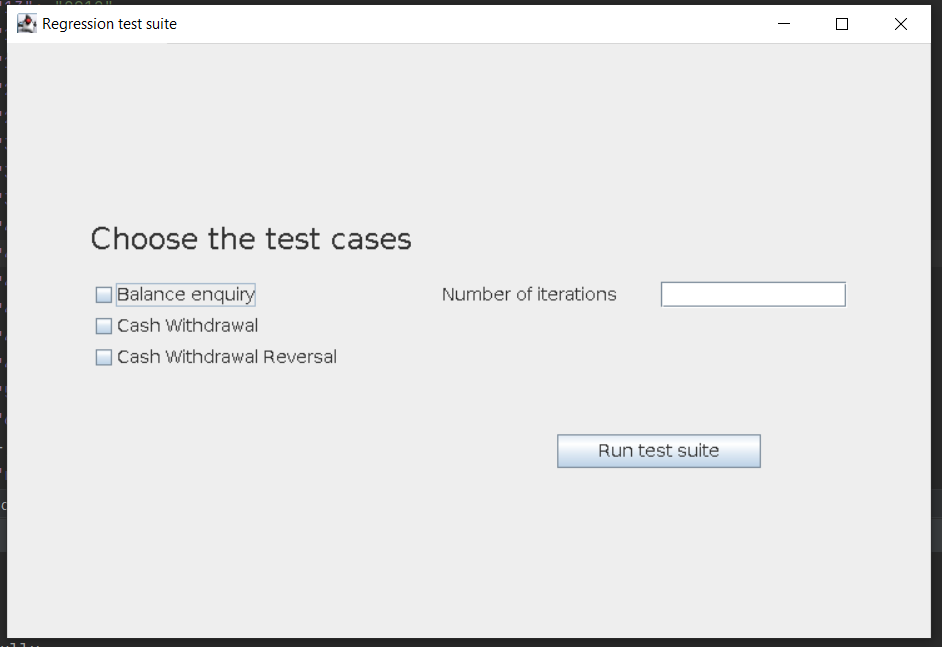}
    \caption{Client GUI}
    \label{fig:my_label}
\end{figure}

\begin{figure}
    \centering
    \includegraphics[scale = 0.5]{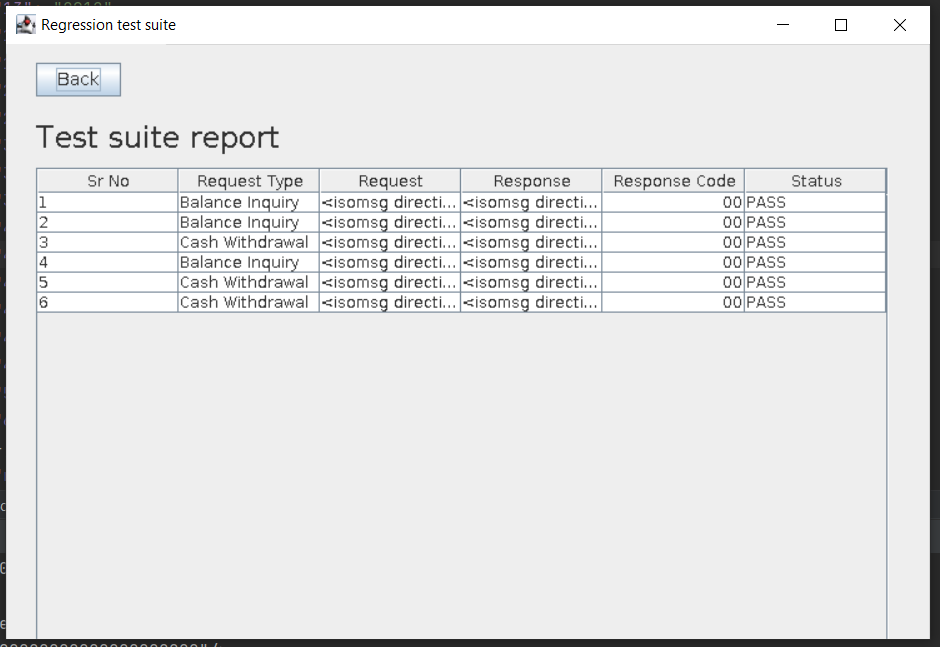}
    \caption{Report Generation GUI}
    \label{fig:my_label}
\end{figure}

\section{Conclusion}
\par Our proposed system can be used to extensively test both key roles of a payment switch, viz. when the bank is an issuer bank as well as when the bank is the acquirer. This was our key objective and we're pleased to have met it. 

\par Given that we send messages asynchronously, we have massively sped up the process of testing. This improves switch developer productivity as they no longer need to spend hours upon hours manually testing the switch for potential bugs. All in all, we're excited to present a system we built that can test such an integral component of payment systems.  

\bibliographystyle{IEEEtran}
\bibliography{references}
\newpage

\end{document}